\documentclass[11pt]{article}
\usepackage[utf8]{inputenc}
\usepackage{amssymb}
\usepackage{setspace}
\usepackage{graphicx}
\usepackage{hyperref}
\usepackage{harvard}

\def\urltilda{\kern -.15em\lower .7ex\hbox{\~{}}\kern .04em}

\long\def\symbolfootnote[#1]#2{\begingroup%
	\def\thefootnote{\fnsymbol{footnote}}\footnote[#1]{#2}\endgroup}

\begin{document}
	
% Simple Diagnostics for Difference-in-Differences Estimation

%%%%%%%%%%%%%%%%%%%%%%%%%%%%%%%%%%%%%% TITLE PAGE %%%%%%%%%%%%%%%%%%%%%%%%%%%%%%%%%%%%%%

\begin{titlepage}
	
	\begin{center}
		\LARGE{\textbf{Simple Diagnostics for Two-Way Fixed Effects}}
	\end{center}
	
	\vspace{0.4in}	
	
	\begin{center}
		\begin{Large}
			Pamela Jakiela\symbolfootnote[1]{%
				Jakiela:  Williams College, BREAD, CGD, IZA, and J-PAL, email: pamela.jakiela@williams.edu.
				I am grateful to Jessica Goldberg and Owen Ozier for helpful comments.  All errors are my own.}
			
		\end{Large}	
	\end{center}

	\begin{center}
		\begin{large}
			\today
		\end{large}
	\end{center}

	\vspace{0.8in}
	
	\begin{abstract}
		
		\noindent
		Difference-in-differences estimation is a widely used method of program 
evaluation. When treatment is implemented in different places at different times, researchers often use two-way fixed effects to control for location-specific and period-specific shocks.  Such estimates can be severely biased when treatment effects change over time within treated units.  I review the sources of this bias and propose several simple diagnostics for assessing its likely severity.  I illustrate these tools through a case study of free primary education in Sub-Saharan Africa.

		\medskip
		

		\medskip
		
	\end{abstract}
	
\end{titlepage}

\addtocounter{page}{1}

\doublespacing

\section{Introduction}

Difference-in-differences estimation is a widely used method of program 
evaluation.  Because it does not require explicit knowledge of the 
rule governing treatment assignment, it can be used for retrospective evaluation in a wide range of settings where pre-treatment data is available.  When a treatment of interest is implemented 
in different places at different times, researchers often use two-way fixed effects to control for location-specific and period-specific shocks, estimating an average treatment effect across locations and time periods.  Recent research demonstrates that such estimates can be severely biased -- and may even be incorrectly signed -- when treatment effects change over time within treated units \cite{GoodmanBacon2020,Chaisemartin2020}. 

I review the sources of this bias, and propose several simple diagnostics for assessing its likely severity.  
When a common trends assumption is satisfied, the two-way fixed effects estimator is a linear combination of the treatment effects across treated units; however, when most or all units are treated in later periods, some treated observations may receive negative weight in the calculation of the estimated treatment effect. 
Negative weights are a natural consequence of the two-way fixed effects specification, and are not in and of themselves a cause for concern.  However, they highlight the extent to which the two-way fixed effects difference-in-differences estimation is not ``model free'' when treatment timing is staggered.\footnote{\citeasnoun{CardKrueger1995} argue that the benefit of natural experiments is that they allow for ``model free'' evaluation of policy impacts (Card and Krueger 1995, p.~24).  However, as recent analysis of difference-in-differences illustrates, the assumption that analysis of natural experiments is model free is not necessarily justified once one moves away from simple comparisons of means between treated and untreated groups.  See \citeasnoun{GibbonsEtAl2019} for a general discussion of mis-specification in fixed effects estimation.}  Indeed, the model reflects a specific set of assumptions about the structural relationship between outcomes and treatment, and mis-specification can lead to severely biased estimates.  

Though negative weighting is appropriate when treatment effects are homogeneous and the two-way fixed effects model is correctly specified, two-way fixed effects does not necessarily yield an unbiased estimate of any weighted average treatment effect when treated units receive negative weight and treatment effects are heterogeneous across units \cite{Chaisemartin2020,GoodmanBacon2020,CallawaySantanna2020,SunAbraham2020,Baker2021}. For example, if the treatment effect were zero for all units except one that received negative weight in two-way fixed effect estimation, the expected sign of the two-way fixed effects coefficient would be opposite that of any weighted average treatment effect.  In general, the two-way fixed effects estimate of the treatment effect need not fall between the minimum and maximum treatment effects on treated units.

I use a simple example -- the elimination of primary school fees in 15 African countries -- to illustrate a set of robustness checks that can be used to address concerns about bias and identify settings where it is unlikely to be a cause for concern.  Simple tests can assess the extent to which a two-way fixed effects estimator places negative weight on treated observations (because the weights are proportional to the residuals from a regression of treatment on the fixed effects), and whether the treatment effect homogeneity assumption required for such an estimator to be unbiased is likely to be appropriate (because treatment effect homogeneity implies a linear relationship between residualized outcomes and residualized treatment after removing the fixed effects).  In many settings, the assumptions necessary for two-way fixed effects to provide an unbiased estimate of a weighted average treatment effect may be satisfied, and robustness checks that omit later observations can address concerns about the potential for bias.

\section{The Two-Way Fixed Effects Estimator}

%We are interested in estimating the impact of eliminating school fees on 
%enrollment in primary school.  Let $D_{it}$ be an indicator equal to one 
%if country $i$ offers free primary schooling in year $t$, and let 
%$Y_{it}$ be gross primary enrollment in country $i$ in year $t$.  
I am interested in estimating the impact of treatment $D_{it}$ on 
outcome $Y_{it}$, where $i$ denotes a geographic unit of observation (e.g.~country) 
and $t$ indicates a time period (e.g.~year).  I will refer to 
units $i$ as countries and time periods $t$ as years throughout.  
Treatment $D_{it}$ varies at the country-year level; once treatment 
starts in country $i$, it remains 
``on'' in all subsequent periods for that country:  if $D_{it} = 1$, then 
$D_{i \tau} = 1$ for all $\tau > t$.  

I am estimating the treatment effect of $D_{it}$ on $Y_{it}$ 
via two-way fixed effects using the regression specification:  
\begin{equation} \label{eq:twfe}
Y_{it} = \lambda_i + \gamma_t + \beta D_{it} + \epsilon_{it}
\end{equation}
where $\lambda_i$ is a vector of country fixed effects and $\gamma_t$ 
is a vector of year fixed effects.  
By applying the Frisch-Waugh-Lovell theorem, we can write the OLS estimate of the treatment effect, $\beta^{\textit{\scriptsize{twfe}}}$, as
\begin{equation} \label{eq:fwl}
\beta^{\textit{\scriptsize{twfe}}} = \sum_{it} Y_{it} \left( \frac{\tilde{D}_{it}}{\sum_{it} \tilde{D}^2_{it} } \right)
\end{equation}
where $\tilde{D}_{it}$ is the residual from a regression of the treatment indicator, $D_{it}$, 
on the country and year fixed effects \cite{Chaisemartin2020}.  In a balanced panel, 
\begin{equation} \label{eq:dtilde}
\tilde{D}_{it} = D_{it} - \bar{D}_t - \bar{D}_i + \bar{D}_{all}
\end{equation}
where $\bar{D}_t$ is the average level of treatment across all observations in year $t$, 
$\bar{D}_i$ is the average level of treatment across all observations for country $i$, and
$\bar{D}_{all}$ is the average level of treatment across the entire sample of country-years \cite{GoodmanBacon2020}.

Thus, $\beta^{\textit{\scriptsize{twfe}}} $ is a weighted sum of the values of the outcome variable across 
all observations in the data set.  When a common trends 
assumption holds such that country-level pre-treatment means and year-level shocks are effectively differenced out 
by the fixed effects, $\beta^{\textit{\scriptsize{twfe}}}$ is, in expectation, a linear combination of the treatment effects 
across country-year observations where $D_{it} = 1$.  Importantly, some treated units may receive negative 
weight, and not all country-years are weighted equally.  Intuitively, this occurs because the inclusion  
of two-way fixed effects transforms the binary treatment indicator $D_{it}$ into a continuous measure of 
treatment intensity not explained by the fixed effects, $\tilde{D}_{it}$.  As in any univariate OLS regression of an 
outcome on a continuous measure of treatment intensity, observations with below mean treatment intensity receive negative weight, and 
may be thought of as part of the comparison group.  However, in the case of two-way fixed effects, it is 
outcomes with below mean levels of \emph{residualized} treatment intensity -- after controlling for country and year 
fixed effects -- that receive negative weight.  

When negative weights occur among observations in the treatment group (i.e.~country-years with $D_{it}=1$), 
they will tend to occur in early-adopter countries (where the country-level treatment mean is high) and 
in later years (when the year-level treatment mean is also high).\footnote{As \citeasnoun{GoodmanBacon2020} demonstrates, 
when treatment timing varies across units that eventually receive treatment, 
$\beta^{\textit{\scriptsize{twfe}}} $ can also be decomposed into a weighted average of all possible 
pairwise 2$\times$2 difference-in-differences estimators that can be constructed from the data.  $\beta^{\textit{\scriptsize{twfe}}} $ is a weighted average 
(with weights summing to one) of three types of 2$\times$2 difference-in-differences estimators:  
(1) comparisons of early adopters with later adopters over periods when the later adopters are not yet treated, 
(2) comparisons of early adopters with later adopters over the periods when the early adopters are already treated 
\emph{using the early adopters as the comparison group}, and (3) comparisons of ever-treated groups 
with the never-treated group, if there is one.  The weight placed on an individual country-year observation 
in calculating $\beta^{\textit{\scriptsize{twfe}}} $ is the sum of the weights 
it receives across all three types of comparisons, including those where it is used as a comparison group.  
While untreated country-years are never used as the treatment group in such pairwise comparisons, 
treated country-years are used a comparison group some of the time.  Country-years that receive negative weight 
in the calculation of $\beta^{\textit{\scriptsize{twfe}}} $ are those that are used primarily as the comparison group 
in the construction of pairwise 2$\times$2 difference-in-differences estimators.}  
Treated country-years that receive negative weight 
are those where the level of treatment predicted by the country and year fixed effects exceeds one.\footnote{Hence, 
there is a parallel with linear probability models, where predicted probabilities outside the unit interval 
may be viewed as an indication of mis-specification.}  Hence, 
a sufficiently large never-treated group combined with enough pre-treatment data will guarantee that 
negative weights do not occur in the treatment group.  However, in data sets with a limited number of 
pre-treatment periods, or with periods in which all or most units are treated, two-way fixed effects estimation 
will often put negative weight on the treatment effects in later periods for early-adopter units.

Negative weights on treated country-year observations are not, by themselves, a cause for concern.  When treatment effects are 
homogeneous, the two-way fixed effects model is correctly specified because the dose response relationship 
between the residualized outcome variable $\tilde{Y}_{it}$ and the residualized treatment variable $\tilde{D}_{it}$ is linear.  
OLS correctly adjusts for the fact that the estimated fixed effects associated with high-treatment units and high-treatment periods are capturing some of the true treatment effect.  Hence, negative weights are not a pathology, but a desired and natural consequence of the (implicit) modeling assumption we make when we difference out country and year means and estimate a single (implicitly homogeneous) treatment effect.

%Negative weights arise, mechanically, because the predicted level of treatment 
%(from a regression of actual treatment status on the country and year fixed effects) in some treated units is greater than one.  
%By including two-way fixed effects in our difference-in-differences specification, we are transforming a regression on a dummy variable (treatment) into a regression on a continuous variable (residualized treatment).  The residual from our regression of treatment on our fixed effects provides a continuous measure of treatment intensity \emph{beyond what is predicted by the fixed effects alone}.  Whenever we estimate a treatment effect using a continuous measure of treatment intensity, we impose a linear dose-response relationship; mis-specification of the dose response relationship can lead to bias in cases where the identification strategy is valid. 

Though negative weights are not a cause for concern when treatment effects are homogeneous, the two-way fixed effects 
estimator can be severely biased when treatment effects are heterogeneous --- particularly when they change over time 
within treated units \cite{Chaisemartin2020,GoodmanBacon2020}.  In these cases, $\beta^{\textit{\scriptsize{twfe}}} $ 
will not necessarily fall between the minimum and maximum treatment effects on any individual country-years.\footnote{This is true  
even in cases where no treated observations receive negative weight, as 
the two-way fixed effects coefficient is effectively re-scaled to fit 
the linear relationship between outcomes and residualized treatment.  However, 
when treatment effects are heterogeneous but are all of the same sign, the expected value of the two-way fixed effects estimator can only be signed incorrectly if treated observations receive negative weight in calculating the estimated effect.  }
For example, \citeasnoun{Baker2021} show that estimates 
of the impact of banking deregulation on inequality in the United States are biased because the impacts of deregulation 
appear to grow larger over time. Hence, it is important to test whether difference-in-differences estimates derived from 
two-way fixed effects estimation are influenced by the inclusion of later country-years receiving negative weight in 
the calculation of the average treatment effect, and -- if so -- whether the assumption of treatment effect homogeneity 
is plausible.

\section{Simple Diagnostics for Two-Way Fixed Effects}

%The foregoing discussion suggests two simple tests that can asses the potential for bias in two-way fixed effects.  
When assessing whether difference-in-differences estimates 
derived from two-way fixed effects estimation are likely to be biased, it is important to answer 
two questions.  First, do any treated units receive negative weight in the calculation of 
$ \beta^{\textit{\scriptsize{twfe}}}$?  Answering this question is straightforward since the 
weights are proportional to the residuals from a regression of treatment on country and year 
fixed effects.  Second, can one reject the hypothesis 
that treatment effects are homogeneous?  To answer this second question, one can exploit the fact that, 
under the assumptions of treatment effect homogeneity and common trends, the relationship between $\tilde{Y}_{it}$ 
and $\tilde{D}_{it}$ is linear; a testable implication is that the slope does not differ between the treatment group and the control group.  

In what follows, I illustrate how these diagnostics can be used in practice.  I present an empirical example, 
assessing the impact of the elimination of primary school fees in Sub-Saharan Africa on school enrollment.    
I use data on gross enrollment in primary and secondary school from 15 African countries 
that eliminated primary school fees between 1994 and 2013.  These outcomes are ideal for illustrating 
the properties of the two-way fixed effects estimator because the elimination of primary school fees is likely 
to have had a large and immediate impact on primary school enrollment, but only a delayed effect on enrollment in secondary 
school -- suggesting that the risk of bias in two-way fixed effects is larger for the latter outcome.  

Data on enrollment comes 
from the World Bank's World Development Indicators database.  My analysis includes data from 1981 through 2015.  
Data on the timing of the elimination of school fees  comes from Koski et al.~(2018).  The countries included 
in the sample and the years that each country eliminated 
primary school tuition fees are listed in Online Appendix Table 1.  The data set contains 15 countries, 
but only 13 distinct timing groups (defined as countries that eliminated school fees in the same year) since 
Kenya and Rwanda both eliminated primary school fees in 2003, while Benin and Lesotho both eliminated fees in 2006.  
All data and code used in this paper is available at \url{https://pjakiela.github.io/TWFE}.

In Online Appendix Table \ref{tab:fpe-effect}, I estimate the impact of eliminating school fees on enrollment in primary and secondary school 
while controlling for country and year fixed effects.  Estimates suggest that introducing free primary education increased 
gross enrollment in primary school by 20 percentage points (p-value 0.04).  These results are consistent with existing 
evidence from specific implementing countries demonstrating that the elimination of schools fees increased enrollment 
(cf.~Lucas and Mbiti 2012, Njeru et al.~2014).  Using the two-way fixed effects specfication, I do not 
find evidence that free primary education led to an increase 
in enrollment in secondary school (estimated coefficient $-0.47$ percentage points, p-value 0.88).

\nocite{LucasMbiti2012}
\nocite{NjeruEtAl2014}
\nocite{Koski2018}

\subsection{Do Treated Observations Receive Negative Weight?}

\begin{figure}
\begin{center}	
\caption{Two-Way Fixed Effects Weights, by Treatment Status} \label{fig:weights}

\medskip
\medskip

Panel A:  Dependent Variable:  Gross Enrollment in Primary School

\includegraphics[width=0.72\textwidth]{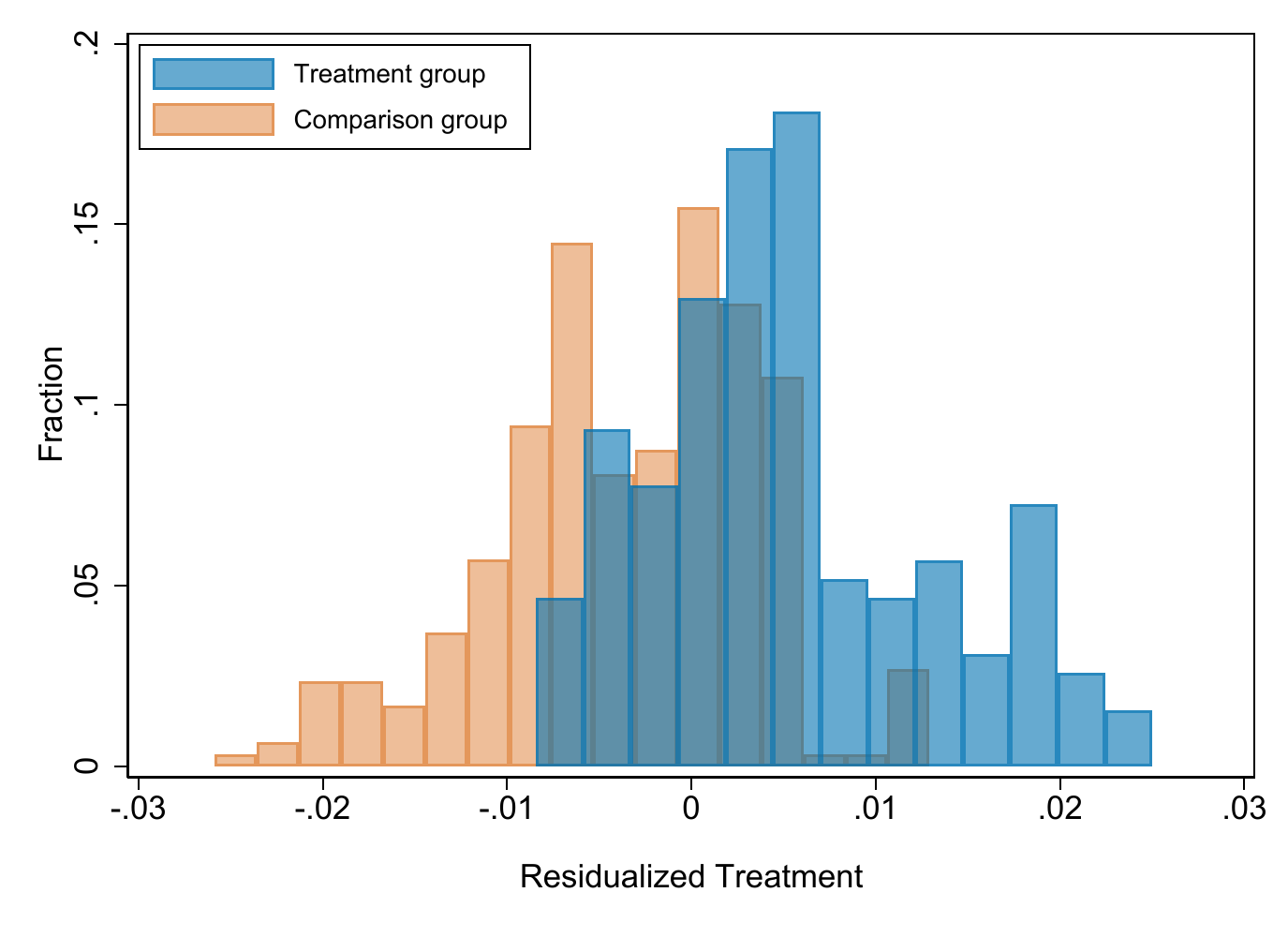}

\medskip

Panel B:  Dependent Variable:  Gross Enrollment in Secondary School

\includegraphics[width=0.72\textwidth]{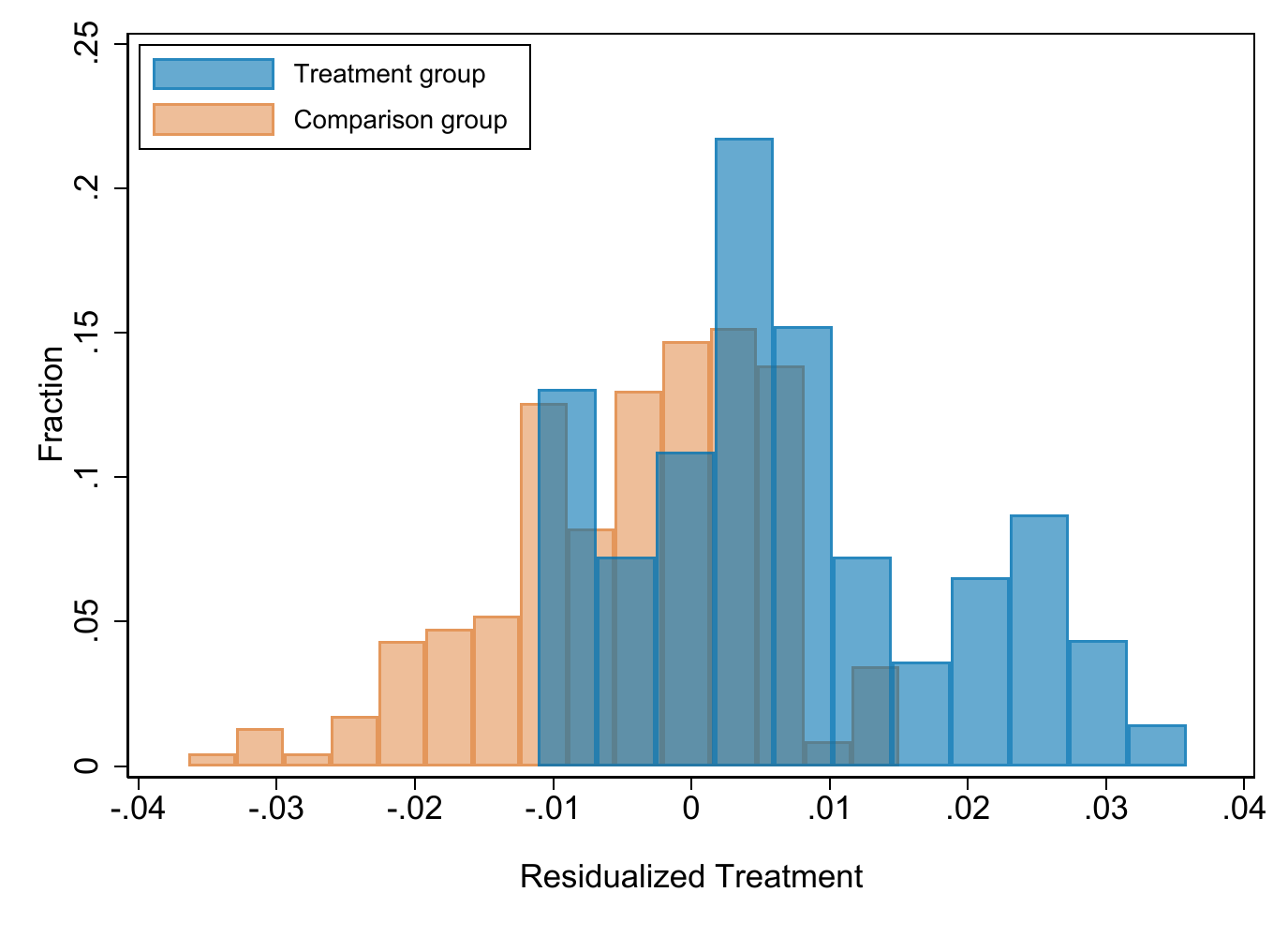}

\end{center}

\footnotesize{The figure presents histograms of the weights used to calculate the two-way fixed effects estimates of the impact of eliminated primary school fees on gross enrollment in primary school (Panel A) and secondary school (Panel B).  The weights are the residuals from a regression of treatment on country and year fixed effects, scaled by the sum of the squared residuals across all observations.  See \citeasnoun{Chaisemartin2020} for discussion.  The weights are not identical in the two specifications because both panels are imbalanced, but the missing country-years differ across the two outcome variables.}
\end{figure}

Figure \ref{fig:weights} plots the weights placed on country-year level observations in calculating 
the two-way fixed effects estimate of the treatment effect of eliminating primary school fees.  As discussed above, 
these weights are proportional to the residuals from a regression of the treatment dummy on the set of country and year fixed effects.  
As expected, the weights sum to zero (across the treatment and control observations), but some \emph{treated} country-year observations 
receive negative weight, and some untreated country-year observations receive positive weight -- so the weights on 
country-years in the treatment group do not necessarily sum to one.  
Approximately 26 percent of all treated country-year observations receive negative weight in 
the estimation of the treatment effect.\footnote{Because data availability differs 
across country-years for the two outcome variables, the residualized treatment variable $D_{it}$ is not identical 
in the two specifications (though the treatment variable is the same in both cases).  In the analysis of impacts 
on primary school enrollment 50 out of 193 non-missing treated country-year 
observations receive negative weight in the calculation of the treatment effect.  
In the analysis of impacts on secondary school enrollment 36 out of 138 non-missing treated country-year observations 
receive negative weight in the calculation of the treatment effect.} 

\begin{figure}
\begin{center}	
\caption{Weights Used in Two-Way Fixed Effects, by Country and Year} \label{fig:heatmap}

\medskip
\medskip

Panel A:  Dependent Variable:  Gross Enrollment in Primary School

\includegraphics[trim=0cm 0cm 0cm 0.75cm,clip,width=0.72\textwidth]{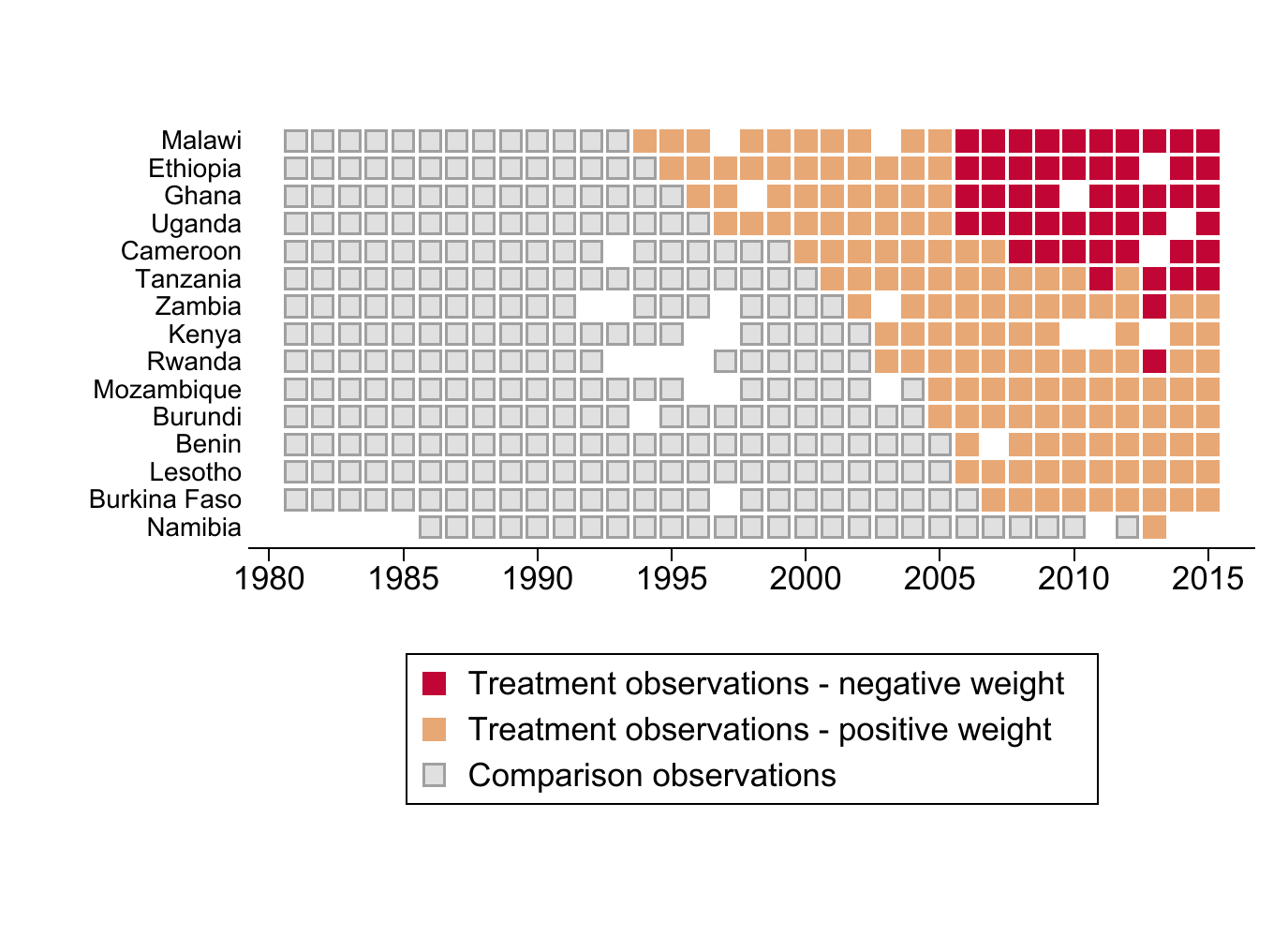}

\medskip

Panel B:  Dependent Variable:  Gross Enrollment in Secondary School

\includegraphics[trim=0cm 0cm 0cm 0.75cm,clip,width=0.72\textwidth]{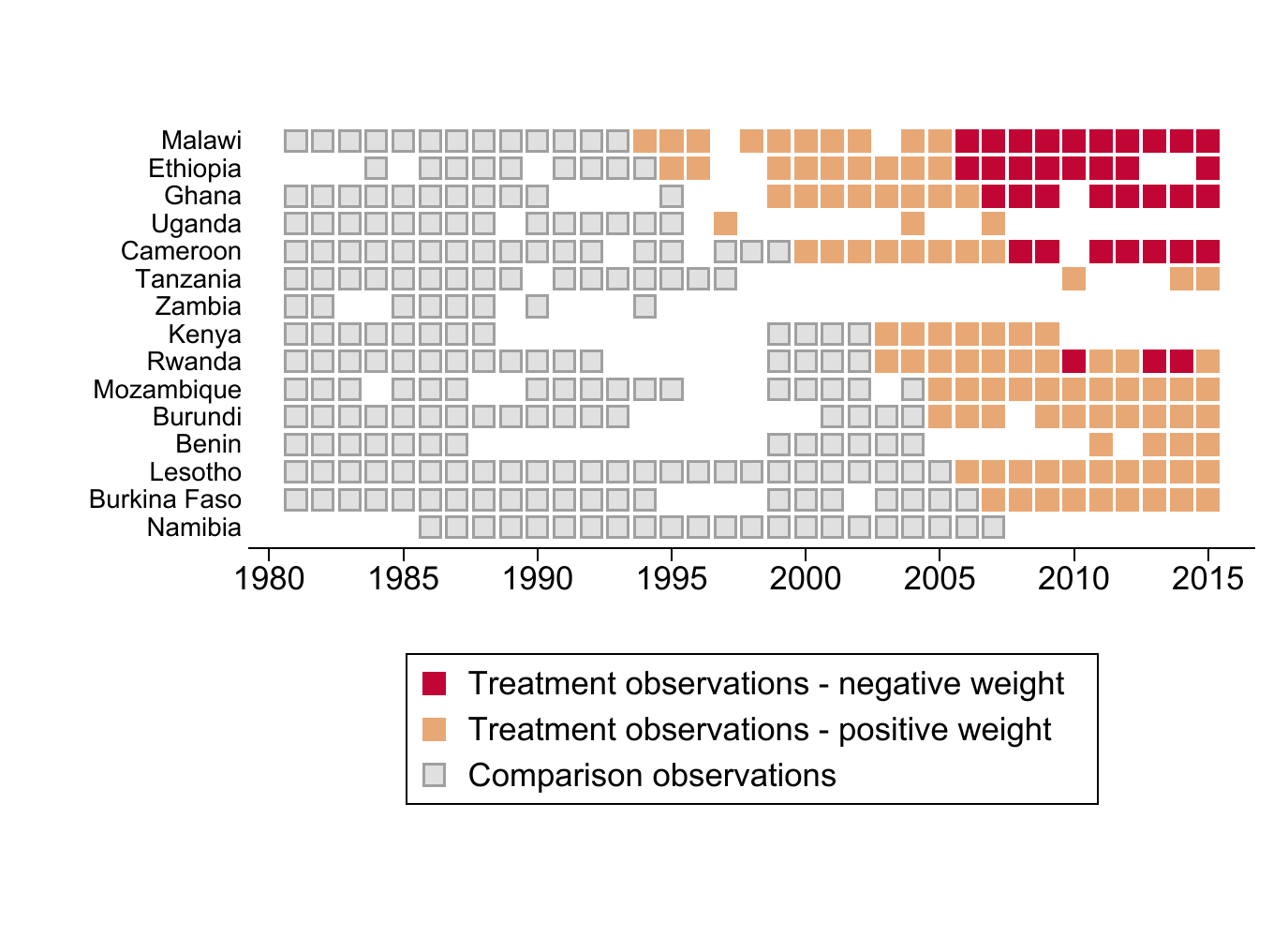}

\end{center}

\footnotesize{The figure characterizes the weights used to calculate the two-way fixed effects estimates of the impact of eliminated primary school fees on gross enrollment in primary school (Panel A) and secondary school (Panel B).  The weights are the residuals from a regression of treatment on country and year fixed effects, scaled by the sum of the squared residuals across all observations.  See \citeasnoun{Chaisemartin2020} for discussion.  The weights are not identical in the two specifications because both panels are imbalanced, but the missing country-years differ across the two outcome variables (as shown in the figure).}

\end{figure}

Figure \ref{fig:heatmap} illustrates the distribution of negative weights across country-year observations.  The unbalanced 
nature of the panel impacts which treated country-years receive negative weight in the estimation of the treatment effect, 
and a country-year may be negatively weighted when subsequent years of data from the same country are not.  Nevertheless, 
the country-years receiving negative weight tend to be the later years of data from early-adopter countries. 
No treated country-year observation 
from before 2006 receives negative weight, and almost all observations receiving negative weight (44 of 50 negatively weighted country-years in the case of 
primary enrollment and negatively weighted 33 out of 36 country-years in the case of secondary school enrollment) 
are concentrated among the first five countries to implement free primary schooling.  This suggests two simple robustness checks 
that I implement in Section \ref{sec:robustness}: (i) dropping the last years in the data set and (ii) retaining only a fixed number of 
post-treatment years per country.  Either of these approaches can be used to identify sub-samples in which the two-way fixed effects 
estimator does not place negative weight on any treated country-years.  However, even two-way fixed effects estimates 
that do not rely on negative weighting should be treated with caution when treatment effects are heterogeneous.   
%in such sub-samples, the estimator is a weighted average 
%of the impact of treatment across treated country-years (assuming the common trends assumption holds).

\subsection{Testing the Homogeneity Assumption Directly}

We have seen that the two-way fixed effects estimate of the impact of free primary education is a linear combination 
of country-year outcomes, and that in our sample some treated country years receive negative weight in this calculation.  
As discussed above, negatively weighting observations in the treatment group is appropriate if treatment effects 
are homogeneous, but can lead to bias when treatment effects change over time within treated units.  If the homogeneity 
assumption holds, the relationship between the residualized outcome variable $\tilde{Y}_{it}$ and the residualized 
treatment variable $\tilde{D}_{it}$ is linear.  To see this, consider a balanced panel.  Let $\mu_i$ denote the value of the outcome 
variable $Y_{it}$ in country $i$ when $t=1$, and let $\eta_t$ be the change in the outcome variable that would occur between 
period $t-1$ and $t$ in the absence of treatment (which is assumed to be constant across countries under common trends).\footnote{Hence, $\eta_1 = 0$.}  Let 
$\delta$ denote the homogeneous treatment effect.  The value of the outcome variable for country $i$ in year $t$ can be written as 
\begin{equation}
Y_{it} = \mu_i + \sum_{\tau = 1}^t \eta_t + \delta D_{it} 
\end{equation}
and the residualized outcome variable, $\tilde{Y}_{it}$ is equal to $\delta \tilde{D}_{it}$.  Hence, under the assumption of homogeneous 
treatment effects (and common trends),  $\tilde{Y}_{it}$ is a linear function of $\tilde{D}_{it}$, and the slope does not differ between the treatment group and 
the comparison group.

\begin{figure}
	\begin{center}	
		\caption{The Association Between Residualized Outcomes and Residualized Treatment} \label{fig:residscatter}
		
		\medskip
		
		Panel A:  Dependent Variable:  Gross Enrollment in Primary School
		
		\includegraphics[width=0.72\textwidth]{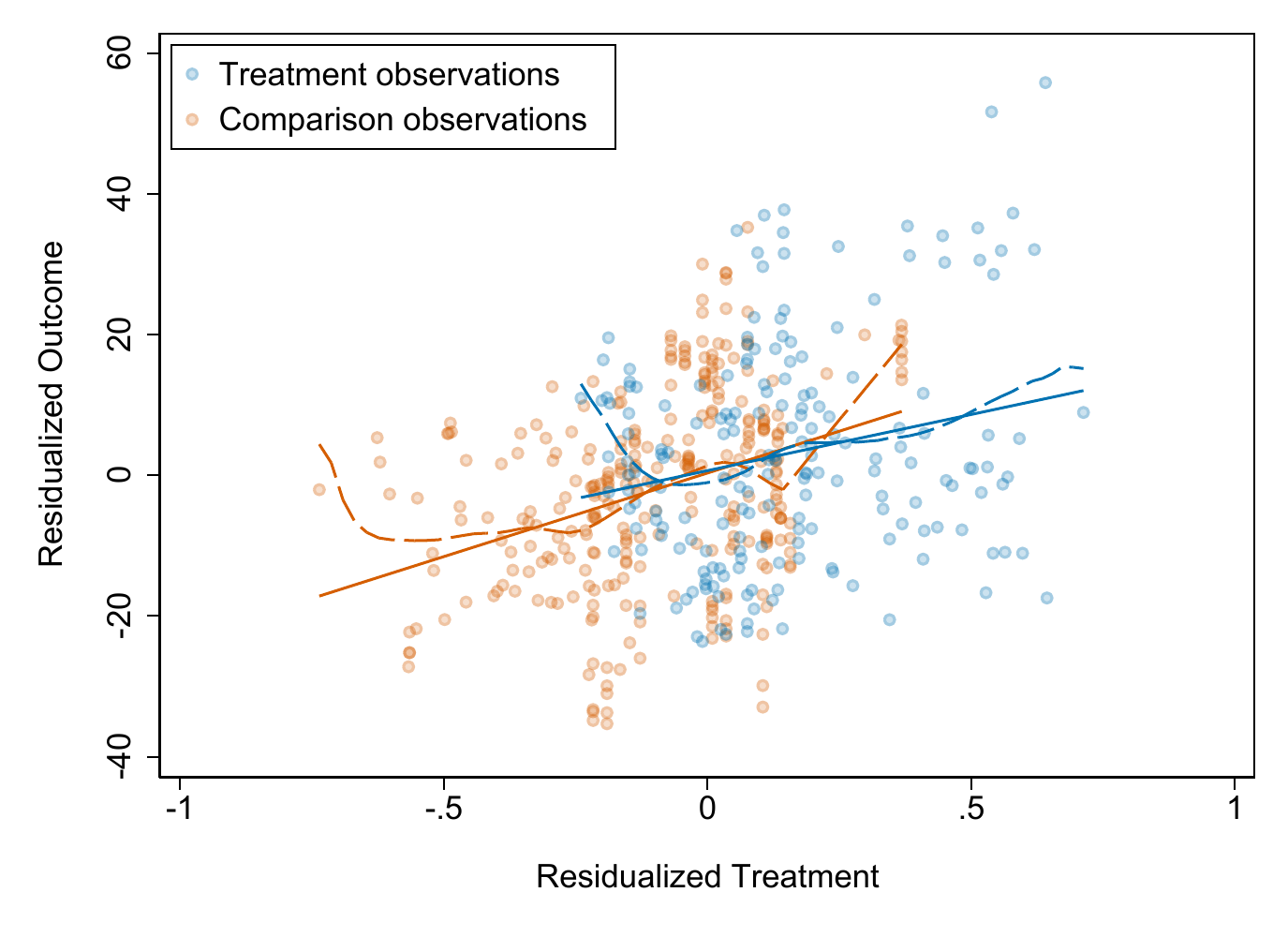}
		
		\medskip
		
		Panel B:  Dependent Variable:  Gross Enrollment in Secondary School
		
		\includegraphics[width=0.72\textwidth]{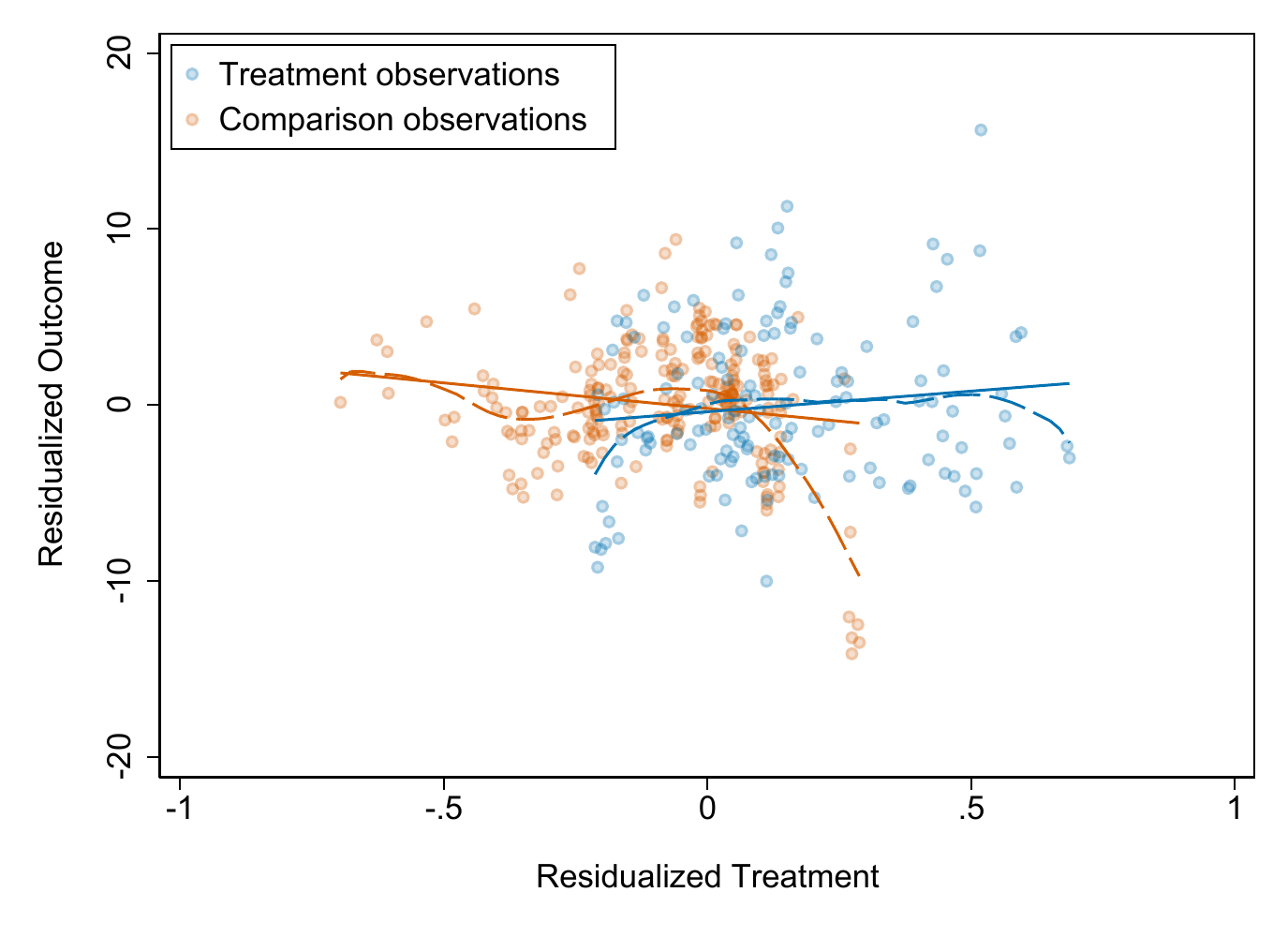}
		
	\end{center}

\footnotesize{The figure plots the relationship between the residuals from a regression of the outcome variable (gross enrollment) on country and year fixed effects and the residuals from a regression of treatment on country and year fixed effects.  If treatment effects are homogeneous, the relationship is linear.  The line of best fit and a local linear regression of residuals from the control group appears in orange; the line of best fit and a local linear regression of residuals from the treatment group appears in blue.}

\end{figure}

Figure \ref{fig:residscatter} presents a scatter plot of these residuals; Panel A shows residuals from the 
regressions using primary school enrollment as the outcome of interest, while Panel B shows the residuals from 
analysis using secondary school enrollment as the outcome.  
In both cases, local polynomial regressions suggest that the relationship between $\tilde{Y}_{it}$ and $\tilde{D}_{it}$ 
may not be perfectly linear, particularly near the extremes of the support.  However, in the case of primary school enrollment (Panel A), 
there is no obvious evidence that the slope differs between the 
treatment group and the comparison group.  OLS regression analysis of $\tilde{Y}_{it}$ on $\tilde{D}_{it}$ confirms 
this (Online Appendix Table \ref{tab:residregs}):  the slope of the estimated linear relationship between $\tilde{Y}_{it}$ and $\tilde{D}_{it}$ 
does not differ between the treatment group and the comparison group (p-value 0.20).  The situation is different when we look at the impacts 
of free primary on secondary school enrollment:  there is clear evidence that the slope of the relationship between the residualized 
outcome variable $\tilde{Y}_{it}$ and the residualized treatment variable $\tilde{D}_{it}$ is not the same for the treatment and comparison group.
Regression analysis confirms this:  the slope of the relationship between $\tilde{Y}_{it}$ and $\tilde{D}_{it}$ is significantly higher for treated country-years than for comparison observations (Online Appendix Table \ref{tab:residregs}, p-value 0.01).  Hence, the relationship between the elimination of 
primary school fees and gross enrollment in secondary school is not consistent with the assumptions required for two-way fixed effects estimation.

\subsection{Robustness Checks} \label{sec:robustness}

%Thus far, we have not seen clear evidence that the assumptions required for two-way fixed effects to provide an unbiased estimate of the impact 
%of free primary education on primary school enrollment are invalid, but we 
%have seen that the assumptions required for valid causal inference through two-way fixed effects estimation do not seem to hold when we examine the impacts of free primary education on secondary school enrollment.  In this section, we further probe the robustness of the estimated treatment effects to assess whether the assumption of homogeneous treatment effects seems plausible.  

Under the assumption of homogeneous treatment effects, dropping some treated country-years from the data set should not affect the expected 
value of the estimated treatment effect (assuming common trends).  Hence, robustness checks of this type, include jackknife estimation, 
can be viewed as further tests of the assumption of treatment effect homogeneity.  I present several examples appropriate to difference-in-differences settings below.

\begin{figure}[h]
	\begin{center}	
		\caption{Robustness to Exclusion of Later Years in Data Set} \label{fig:dropyears}

\medskip		
\medskip

Panel A:  Dependent Variable:  Gross Enrollment in Primary School
		
		\includegraphics[width=0.72\textwidth]{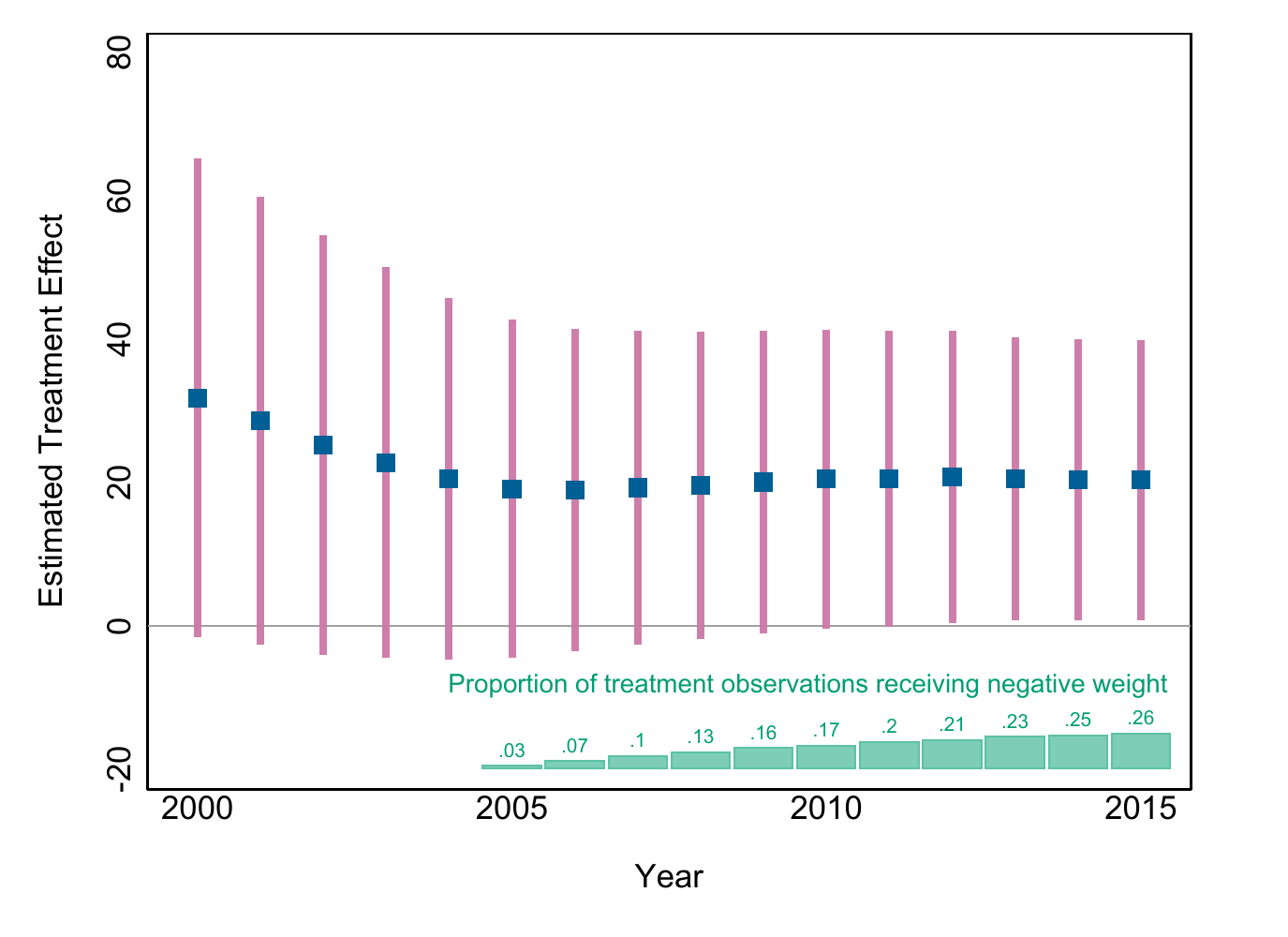}
		
	\medskip
	
	Panel B:  Dependent Variable:  Gross Enrollment in Secondary School
	
	\includegraphics[width=0.72\textwidth]{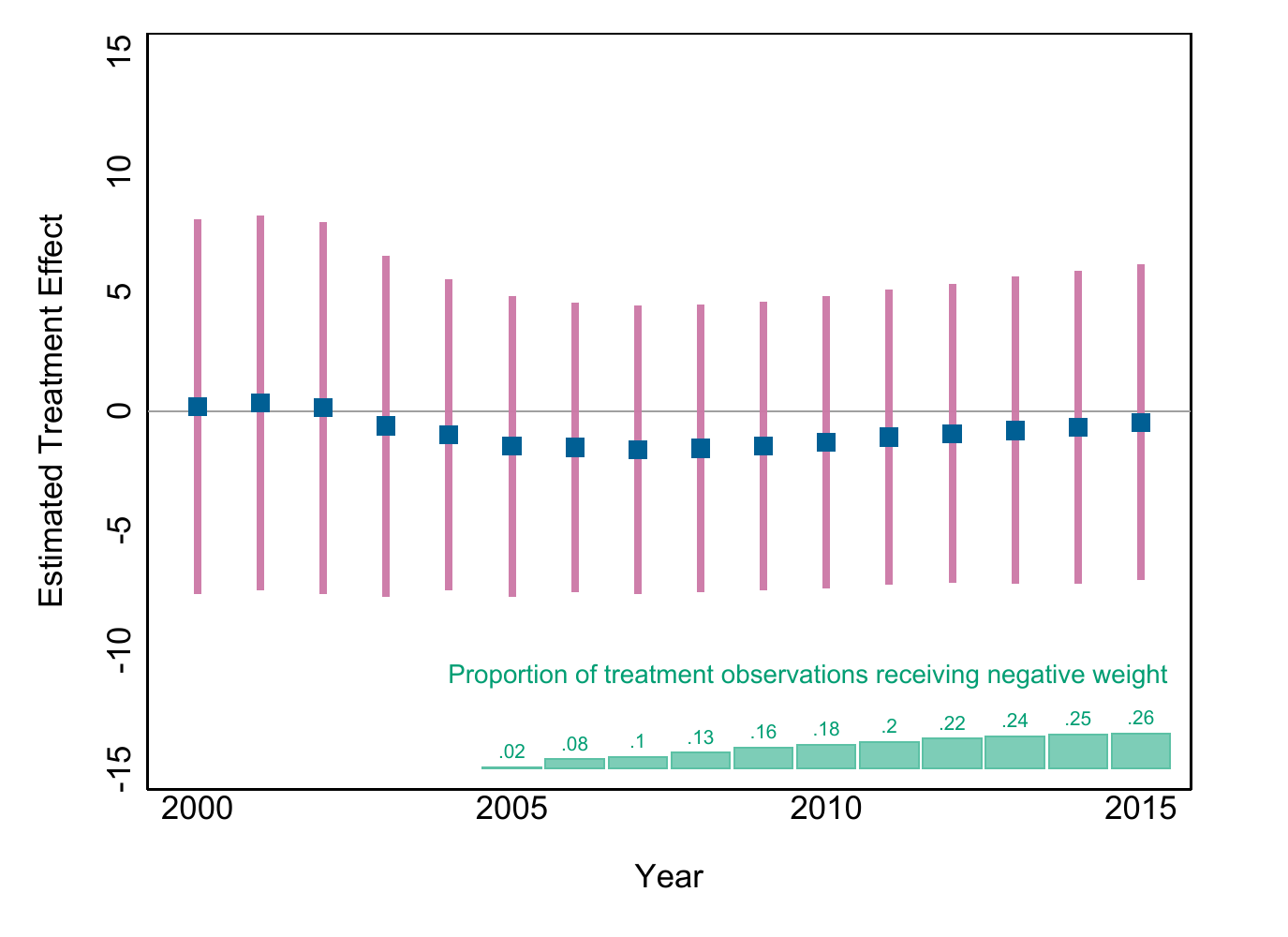}
		
	\end{center}

\footnotesize{The figure summarizes the estimated two-way fixed effects coefficients and the associated confidence intervals from regressions of gross enrollment in primary school (Panel A) and gross enrollment in secondary school (Panel B) on the indicator for free primary education.  Each coefficient represents a different regression that includes data from 1981 through the year indicated on the $x$-axis.  Green bars indicate the proportion of treatment country-years that are negatively weighted in the calculation of the two-way fixed effects coefficient.}
\end{figure}

Figure \ref{fig:dropyears} presents specifications that omit later years from the data set.  For primary school enrollment (Panel A of Figure \ref{fig:dropyears}), 
the estimated treatment effect declines as the last year included in the analysis increases from 2000 to 2004, though no country-years in the treatment group are negatively weighted in calculating these treatment effect -- suggesting treatment effect heterogeneity either across countries (earlier vs.~later adopters) or over time (in the first few years after treatment vs.~later years, among early-adopter countries).  As the last included year increases from 2005 to 2015, the proportion of treatment negatively weighted country-year observations increases from three percent in 2005 to 26 percent in 2015.  However, the estimated treatment effect remains remarkably stable across these specifications -- suggesting that the negatively weighted observations are not driving the results.  When we examine impacts on secondary school enrollment (Panel B of Figure \ref{fig:dropyears}), the estimated treatment effect appears quite stable across specifications -- though we know that the linear relationship between residualized outcomes and residualized treatment does not hold in this case.

Online Appendix Figure \ref{fig:droptime} presents a related robustness check:  varying the number of years of post-treatment data retained for each country.  In this case, the estimated impact on primary enrollment is quite stable across specifications, but the estimated impact on secondary enrollment increases as more post-treatment years are included in the analysis.  The stability in the estimated impact on primary enrollment suggests that heterogeneity apparent in Figure \ref{fig:dropyears} resulted from unusually large impacts in the first countries to implement free primary education rather than changes in treatment effects over time within each country -- possibly because the first countries to implement free primary education had lower pre-FPE levels of gross enrollment.\footnote{The four countries that implemented free primary education in the 1990s -- Malawi, Ethiopia, Ghana, and Uganda -- had substantially lower gross enrollment in the 1980s (with an average gross enrollment ratio of 61.1) than the 11 countries that implemented free primary in 2000 or later (those countries had an average gross enrollment ratio of 79.3 in the 1980s).}

In Online Appendix Figure \ref{fig:dropcountries}, we implement a final robustness check:  dropping individual countries from the analysis.  Again, there is suggestive evidence of some treatment effect heterogeneity, though we can never reject the equality of the estimated coefficients across specifications.  All specifications suggest that eliminating primary school fees increased gross primary enrollment (Panel A of Online Appendix Figure \ref{fig:dropcountries}), though the estimated treatment effect is slightly smaller when Malawi or Uganda (two of the earliest adopters) are excluded -- consistent with the evidence discussed above.  The estimated impacts on secondary school enrollment appear quite robust unless Malawi (the first country in the data set to eliminate primary school fees) is excluded.  In both cases, the evidence suggests that the assumption of treatment effect heterogeneity should be treated with caution.  

\section{Conclusion}

In this paper, we use a case study of free primary education in Sub-Saharan Africa to illustrate a set of simple diagnostics and robustness checks 
that can be used to assess the potential for bias in two-way fixed effects.  When a common trends assumption holds, the two-way fixed effects estimator is a linear combination of outcomes in treated country-years, but in some cases a subset of treated country-years are negatively weighted.  One can test for the extent of negative weighting, since the weights are proportional to the residuals from a regression of treatment on country and year fixed effects.  

Negative weights are not a problem when treatment effects are homogeneous; in fact, they are a natural consequence of the (implicit) assumption of treatment effect homogeneity in two-way fixed effects.  When the assumption of homogeneous treatment effects is valid, the residuals from a regression of the outcome variable on country and year fixed effects are linearly related to the residuals from a regression of treatment on country and year fixed effects -- and the slope of this linear relationship does not differ between the treatment group and the comparison group.  

When negative weights are present or there is evidence of treatment effect heterogeneity, a range of alternative estimators are available \citeaffixed{CallawaySantanna2020,SunAbraham2020}{cf.}.\footnote{See \citeasnoun{Baker2021} for an accessible introduction to the estimators proposed by \cite{CallawaySantanna2020} and \cite{SunAbraham2020}.}  However, these estimators may have less statistical power than the pooled estimator, 
so it may be preferable to use traditional two-way fixed effects when the underlying identification assumptions seem plausible.  I have proposed a simple 
test of treatment effect heterogeneity, but researchers should be alert to the possibility that the test could lack sufficient power in their data or context.  Statistical reasoning should be combined with introspection in assessing whether the assumptions required for two-way fixed effects estimation seem appropriate to the setting.

Even when no treated country-years receive negative weight, the two-way fixed effects estimator does not place equal weight on all treated country-years and imposes a linear dose-response relationship that may be mis-specified \cite{Baker2021,SunAbraham2020}.  When impacts are heterogeneous, it is not clear that applied researchers will wish to weight early-adopters more heavily than later-adopters -- as would happen if all treated country-years received equal weight.  When treatment effects are not homogeneous, many different policy-relevant average treatment effects may exist.  Researchers may prefer to characterize the extent and nature of heterogeneity -- for example, through jackknife estimation and other forms of sensitivity analysis -- rather than focusing on a single average treatment effect that may or may not be policy relevant or externally valid, depending on the context.  As the range of difference-in-differences related approaches expands, researchers have more opportunity to choose a method that identifies their estimand of interest.

Difference-in-differences estimation through two-way fixed effects is one of the most widely used approaches for evaluating social policy.  Recent research highlights the potential pitfalls of two-way fixed effects, but these criticisms should not be interpreted as an indication that two-way fixed effects is no longer a credible identification strategy.  Instead, these criticisms and the newly-developed tools that have emerged in parallel provide an analytical framework for more nuanced and better understood difference-in-differences estimation via two-way fixed effects.

\newpage
\singlespacing
\bibliographystyle{aer}
\bibliography{twfebib} 

\newpage

\clearpage

\appendix
\renewcommand{\thepage}{\thesection\arabic{page}}
\setcounter{page}{1}
\section{Online Appendix:  not for print publication}
\renewcommand{\thefigure}{\thesection\arabic{figure}}
\renewcommand{\thetable}{\thesection\arabic{table}}
\setcounter{table}{0}
\setcounter{figure}{0}

\doublespacing

\begin{table}[h]
	\begin{center}
		\caption{Countries Included in the Analysis} \label{tab:fpe-years}
		
		\medskip
		
		\begingroup
		\setlength{\tabcolsep}{6pt} % Default value: 6pt
		\renewcommand{\arraystretch}{1.4} % Default value: 1
		\begin{tabular}{ll} 
			 & \\	
			\hline \hline 	
			 \textsc{Country} & \textsc{FPE} \\ 
			\hline
			Benin	& 2006 \\
			Burkina Faso	& 2007 \\
			Burundi 	& 2005 \\
			Cameroon	& 2000 \\
			Ethiopia 	& 1995 \\
			Ghana	& 1996 \\ 
			Kenya	& 2003 \\ 
			Lesotho	& 2006 \\
			Malawi	& 1994 \\
			Mozambique	& 2005 \\
			Namibia	& 2013	\\
			Rwanda	& 2003 \\
			Tanzania	& 2001 \\
			Uganda	& 1997 \\
			Zambia	& 2002 \\
			\hline
			\multicolumn{2}{p{3.6cm}}{\footnotesize{\textsc{FPE} indicates the year free primary education was introduced as national policy.  Data on the introduction of free primary education comes from Koski et al.~(2018).}} \\
		\end{tabular}
		\endgroup
	\end{center}
\end{table}

\begin{table}[h]
\begin{center}
\caption{Difference-in-Differences Estimates of Impacts of Free Primary Education} \label{tab:fpe-effect}

\medskip

\begingroup
\setlength{\tabcolsep}{6pt} % Default value: 6pt
\renewcommand{\arraystretch}{1.4} % Default value: 1
\begin{tabular}{lcc} 
\multicolumn{1}{p{4.0cm}}{} & \multicolumn{1}{p{4.0cm}}{} & \multicolumn{1}{p{4.0cm}}{} \\	
\hline \hline 	
 & \multicolumn{2}{c}{\emph{Dependent Variable:  Gross Enrollment in...}} \\ 
	& \textsc{Primary School} & \textsc{Secondary School} \\ 
	& (1) & (2) \\ 
\hline
Free primary education & 20.43 & -0.47 \\ 
& (9.12) & (3.08)  \\  
& [0.04] & [0.88]\\ 
Country fixed effects & Yes & Yes  \\
Year fixed effects & Yes  & Yes \\
\hline
\multicolumn{3}{p{12.6cm}}{\footnotesize{Dependent variable:  gross enrollment ratio.  Data on gross enrollment ratio in 15 countries comes from the World Development Indicators, years 1981 through 2015.  Standard errors (clustered at the country level) in parentheses; p-values in brackets.}} \\
\end{tabular}
\endgroup
\end{center}
\end{table}

\begin{table}[h]
	\begin{center}
		\caption{Testing Relationship Between Residualized Outcomes, Residualized Treatment} \label{tab:residregs}
		
		\medskip
		
		\begingroup
		\setlength{\tabcolsep}{6pt} % Default value: 6pt
		\renewcommand{\arraystretch}{1.4} % Default value: 1
		\begin{tabular}{lcc} 
			\multicolumn{1}{p{4.0cm}}{} & \multicolumn{1}{p{4.0cm}}{} & \multicolumn{1}{p{4.0cm}}{} \\	
			\hline \hline 	
			& \multicolumn{2}{c}{\emph{TWFE Dep.~Var.:  Gross Enrollment in...}} \\ 
			& \textsc{Primary School} & \textsc{Secondary School} \\ 
			& (1) & (2) \\ 
			\hline
			Residualized treatment & 23.76 & -2.90 \\ 
			& (3.97) & (1.36)  \\  
			& [0.00] & [0.03]\\ 
			Treatment group & 0.34 & -0.19 \\ 
			& (1.51) & (0.47)  \\  
			& [0.82] & [0.69]\\ 
			Treatment group $\times$ residualized treatment & -7.81 & 5.25 \\ 
			& (6.07) & (1.99)  \\  
			& [0.20] & [0.01]\\ 
			\hline
			\multicolumn{3}{p{15.1cm}}{\footnotesize{Dependent variable is the residual from a regression of the gross enrollment ratio on country and year fixed effects.  Residualized treatment is the residual from a regression of the treatment dummy (an indicator for years when primary education was free) on country and year fixed effects.  Data on gross enrollment ratio in 15 countries comes from the World Development Indicators, years 1981 through 2015.  Standard errors in parentheses; p-values in brackets.}} \\
		\end{tabular}
		\endgroup
	\end{center}
\end{table}

\clearpage

\begin{figure}[h]
	\begin{center}	
		\caption{Robustness to Exclusion of Later Post-Treatment Years} \label{fig:droptime}
		
		\medskip		
		\medskip
		
		Panel A:  Dependent Variable:  Gross Enrollment in Primary School
		
		\includegraphics[width=0.72\textwidth]{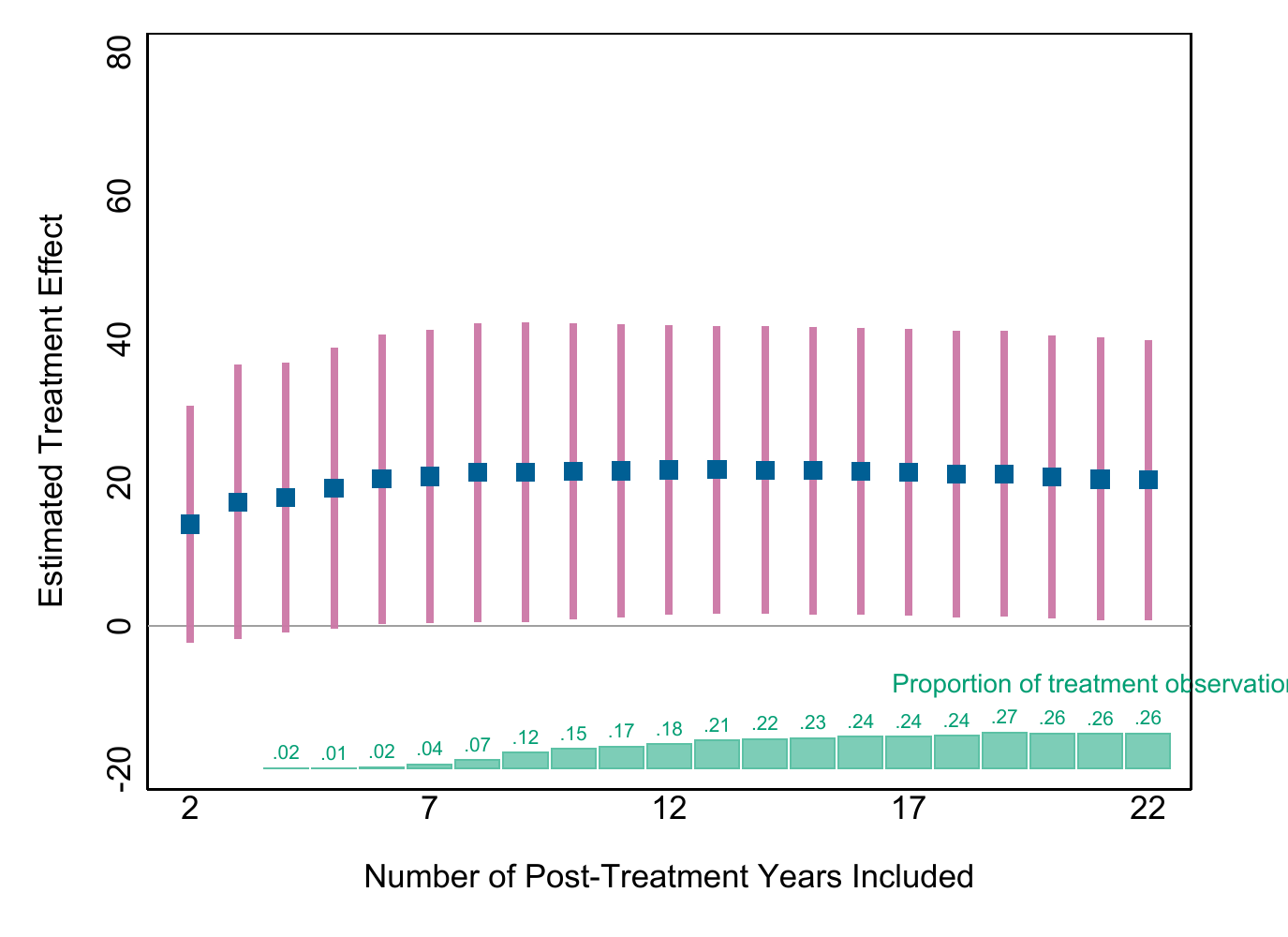}
		
		\medskip
		
		Panel B:  Dependent Variable:  Gross Enrollment in Secondary School
		
		\includegraphics[width=0.72\textwidth]{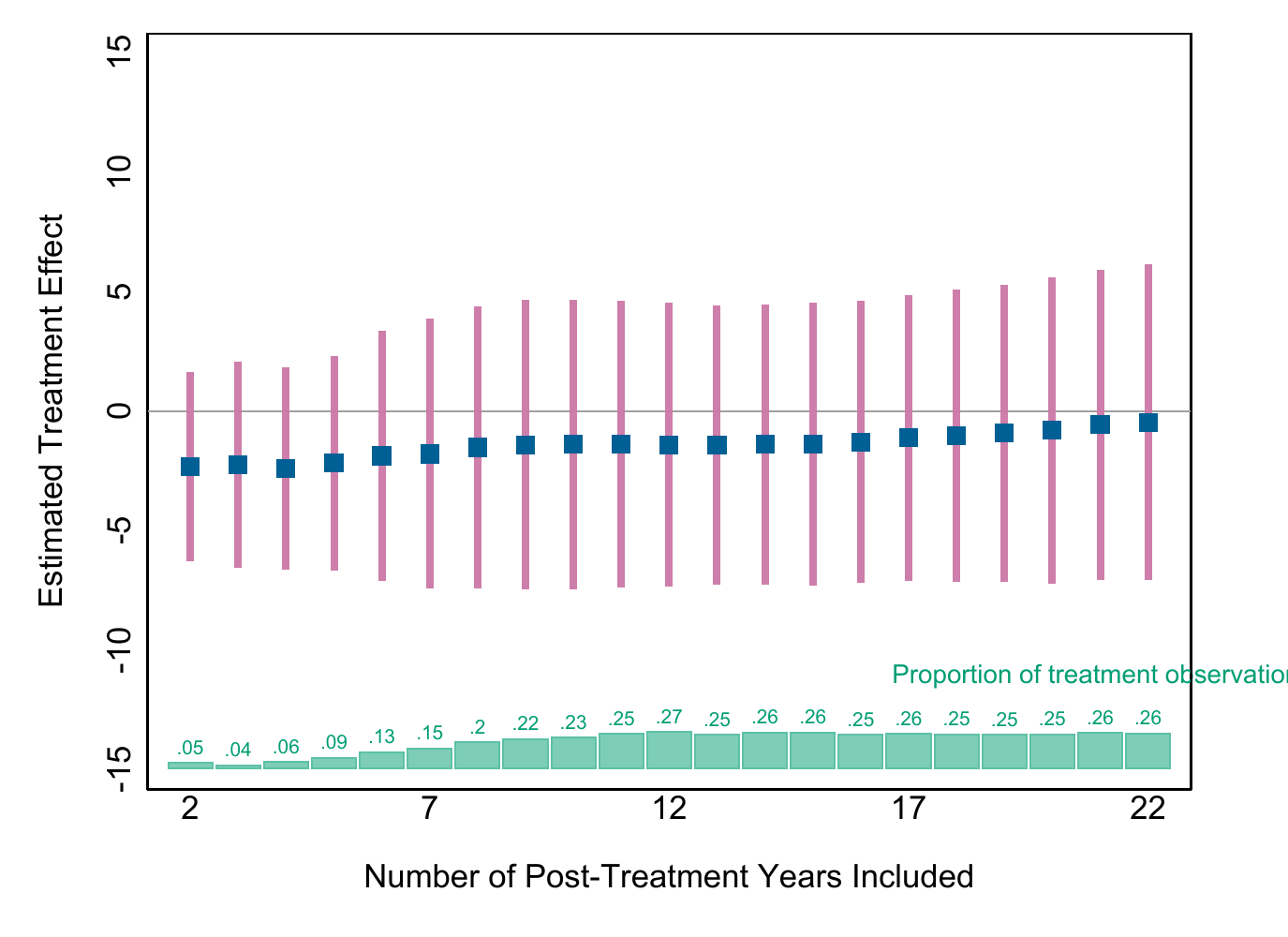}
		
	\end{center}

\footnotesize{The figure summarizes the estimated two-way fixed effects coefficients and the associated confidence intervals from regressions of gross enrollment in primary school (Panel A) and gross enrollment in secondary school (Panel B) on the indicator for free primary education.  Each coefficient represents a different regression that includes data from 1981 through the year $x$ years after the introduction of free primary education (within a given country).  Green bars indicate the proportion of treatment country-years that are negatively weighted in the calculation of the two-way fixed effects coefficient.}
\end{figure}

\clearpage

\begin{figure}[h]
	\begin{center}	
		\caption{Robustness to Exclusion of Individual Countries} \label{fig:dropcountries}
		
		\medskip		
		\medskip
		
		Panel A:  Dependent Variable:  Gross Enrollment in Primary School
		
		\includegraphics[width=0.72\textwidth]{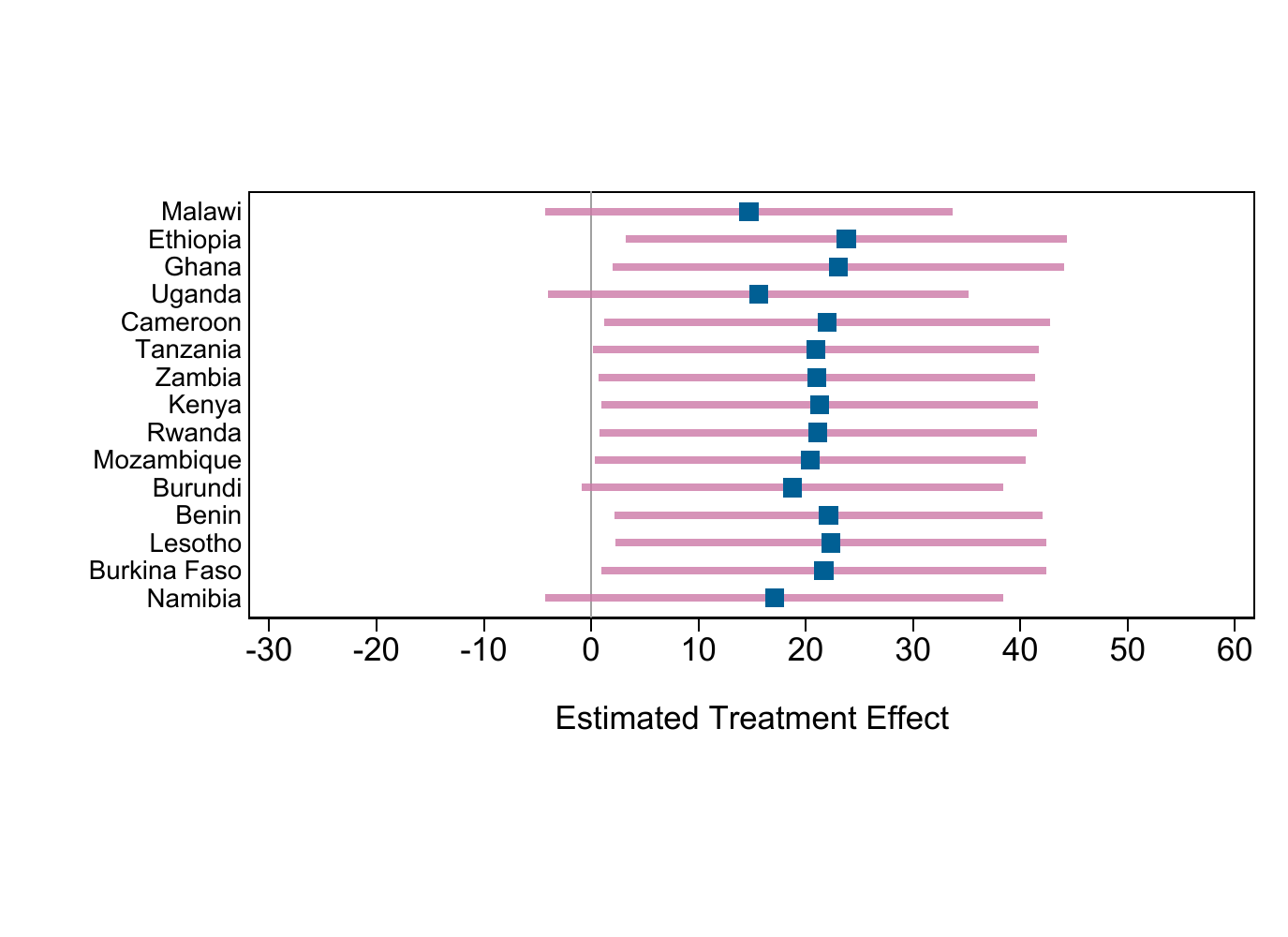}
		
		\medskip
		
		Panel B:  Dependent Variable:  Gross Enrollment in Secondary School
		
		\includegraphics[width=0.72\textwidth]{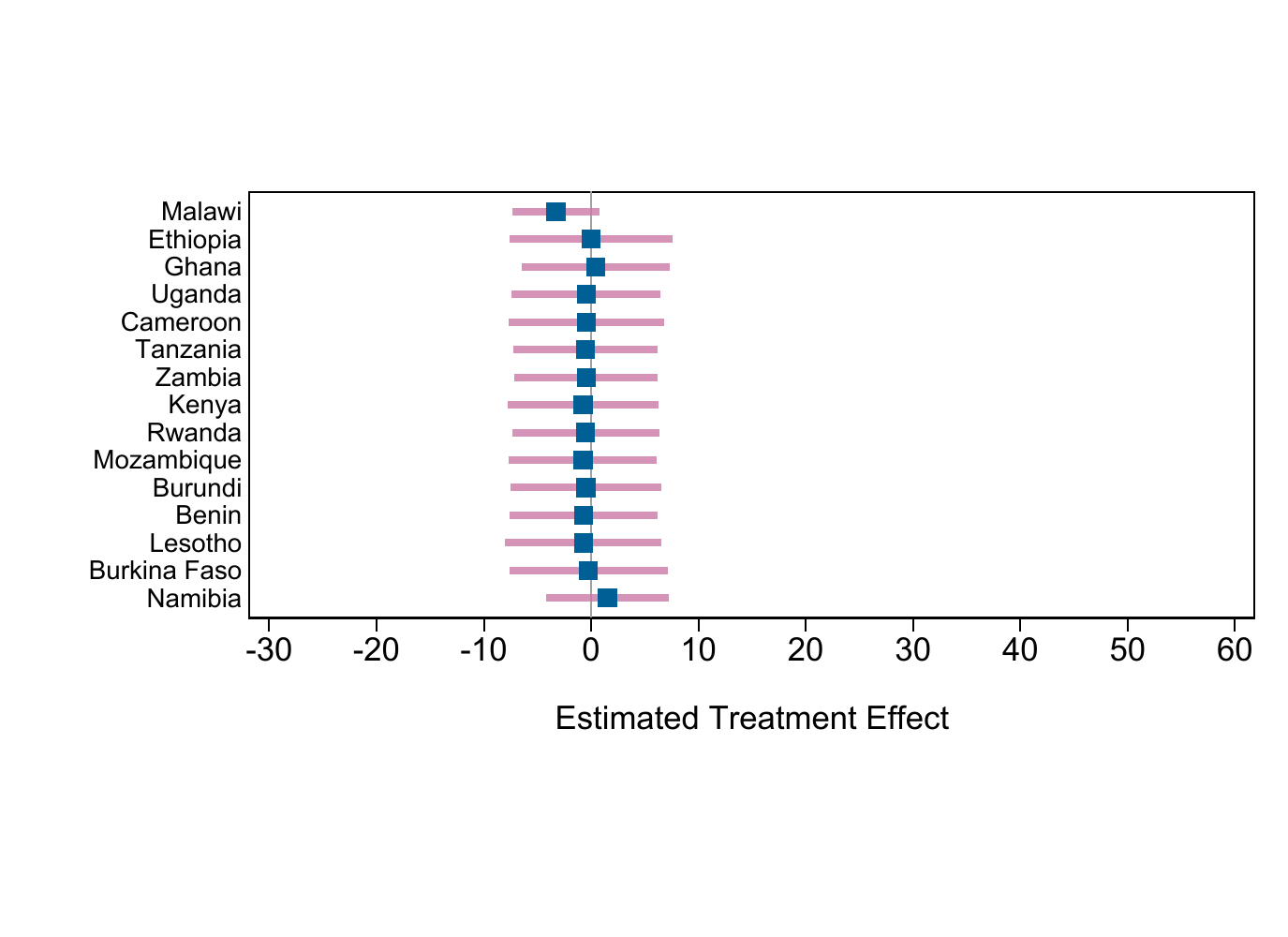}
		
	\end{center}

\footnotesize{The figure summarizes the estimated two-way fixed effects coefficients and the associated confidence intervals from regressions of gross enrollment in primary school (Panel A) and gross enrollment in secondary school (Panel B) on the indicator for free primary education.  Each coefficient represents a different regression that omits the country indicated on the $y$-axis.}

\end{figure}

\end{document}